# Effect of phase and time coupling on NMR relaxation rate by random walk in phase space

Guoxing Lin*

*Carlson School of Chemistry and Biochemistry, Clark University, Worcester, MA 01610, USA*

*Email: glin@clarku.edu

**Abstract**

Phase-time coupling is a natural process in the phase random walks of spin system; however, its effect on the nuclear magnetic resonance (NMR) relaxation is a challenge to the established theories such as the second-order quantum perturbation theory. The recently developed phase diffusion method provides a convenient tool to treat the phase-time coupling effect. From the coupled and uncoupled phase diffusion in the static frame and the rotating frame, the phase diffusion coefficients are obtained, which shows the phase-time coupling has a significant impact on the NMR relaxation rate: The angular frequency $\omega$ in the spectral density is modified to an apparent angular frequency $\eta\omega$, where $\eta$ is the phase-time coupling constant. The strongest coupling has $\eta$ equaling 2, while $\eta$ equaling 1 corresponds to the traditional results. As an example, the modified relaxation time expressions based on both mono-exponential and non-mono-exponential functions can successfully fit the previously reported $^{13}$C $T_1$ NMR experimental data of polyisobutylene (PIB) in the blend of PIB and head-to-head poly(propylene) (hhPP). In contrast, the traditional relaxation rate expression based on the monoexponential time correlation function cannot fit such experimental data. With phase-time coupling, the obtained characteristic time of the segmental motion is faster than that from conventional results.

## I. INTRODUCTION

Nuclear magnetic resonance (NMR) relaxation is a powerful technique for detecting molecular dynamics [1,2,3] in biological or polymer systems [4,5,6]. NMR relaxation is a recovery process in which a spin system's population returns to equilibrium after being perturbed. The molecular thermal motion changes relative molecular orientations, which modulates many fundamental Hamiltonians, including dipolar coupling, quadrupolar coupling, chemical shift anisotropy, etc. [1,2] These modulated Hamiltonians can be viewed as random fluctuating fields exerting on the affected spins. Under the influence of the fluctuating field, the phase evolution of the relevant spins undergoes random walks in phase space, which can be treated by the effective phase diffusion equation [7]. Effective phase diffusion equations have been developed to describe the phase evolution of spin coherence affected by the pulsed-field gradient (PFG) [8] and NMR chemical exchange [9]. The phase diffusion method possesses certain advantages. One of its unique advantages is that the phase distribution in NMR experiments can often be obtained. For example, a well-known PFG approximation is the Gaussian phase distribution (GPD) approximation; in contrast, rather than an approximation, the GPD is an exact solution from the phase diffusion equation method [8]. Additionally, unlike the real space method, the phase diffusion method directly handles the phase evolution process in phase space, which often reduces the degree of solving complexity for analyzing NMR phenomena. For instance, the phase diffusion methods can handle anomalous diffusions in PFG experiments [8], anomalous exchange processes [9], and fractional NMR relaxation [7]. These nonlinear phenomena are often observed in polymer or biological systems, but in general, they are challenges to conventional theories. The phase diffusion method can help us advance in the nonlinear NMR field. This paper is a continuous effort to extend the phase diffusion method to uncover hidden features of NMR relaxations, which should improve our understanding and analysis of the related experiments.

The phase diffusion method offers additional insights into the NMR relaxation study [7]. The impact of the randomly fluctuating field on spin evolution is often treated by the density operator theory based on quantum mechanics [1,2]. Traditional theories such as Bloch-Wangsness-Redfield and the second-order perturbation theories have successfully explained many normal NMR relaxation processes [1,2]. While, Ref. [7] has proposed the phase diffusion equation method, and the phase diffusion coefficient derived in Ref. [7] has the same expression as the relaxation rate given by the conventional theory for normal diffusion; thus, the random phase walk method provides an alternative way to describe NMR relaxation. Compared to traditional theories, the phase diffusion method is intuitive and can conveniently treat complex random walk processes, such as fractional relaxations [7].

Although the fundamental phase diffusion equation method has been built in Ref. [7], the phase and time coupling in the phase random walk has not been investigated. The phase-time coupling connects the phase jump length with its jump time duration in the random walk [10,11,12]. From the coupled continuous time random walk theory, the spatial and temporal coupling has significant effects on the random walk outcome [10,12], which changes the second moment or the variance of the random walk process. The variance of random walk directly affects the diffusion coefficient. Ref. [7] indicates that the phase diffusion coefficient is the same as the relaxation rate that is inversely proportional to the spin-lattice time $T_1$ and spin-spin relaxation time $T_2$ in NMR experiments. Additionally, the coupling between the phase jump length and jump time occurs naturally in each random walk step because the phase shift is directly proportional to the spin moment's precessing time [1,2,7]. Considering the naturally occurring coupling and its compelling impact on the phase diffusion outcome, it is necessary to develop the theoretical treatments for phase-time coupling in the NMR relaxation process.

The phase diffusion method provides a convenient tool to study the phase-time coupling that is difficult to tackle with traditional methods. In this paper, the phase-time coupling effects on the phase diffusion coefficient will be studied in detail, mainly by the continuous time random walk (CTRW) theoretical method.

Both the normal and anomalous phase random walks are investigated in this paper. Anomalous NMR relaxation could arise when anomalous relative particle motion in real space modulates the random field [7], where the fractional rotational and translational diffusion have been proposed to describe the NMR relaxation rate [13,14,15,16]. The fractional rotational and translational diffusions have been applied to other relaxation processes, such as dielectric relaxation [6], relaxation in disordered systems [17], and stress-strain relaxation in viscoelastic materials [18]. Additionally, the relaxation itself, $T_1$ and $T_2$, could be anomalous as described by fractional Bloch-equation proposed by Ref. [19], which gives a Mittag-Leffler function-based NMR relaxation. The monoexponential correlation function is insufficient to describe anomalous relative motion. In a complex system, the time correlation function could either be a Mittag-Leffler function $E_\alpha\left(-\left(\frac{t}{\tau}\right)^\alpha\right)$ [20,21], $E_\alpha(-t^\alpha) = \sum_{n=0}^\infty \frac{(-t^\alpha)^n}{\Gamma(n\alpha+1)}$, or a stretched exponential function $\exp\left(-\left(\frac{t}{\tau}\right)^\alpha\right)$, where $\alpha$ is the order of the time-fractional derivative, and $\tau$ is the characteristic time. The Mittag-Leffler function reduces to a stretched exponential function $\exp\left(-\frac{t^\alpha}{\Gamma(1+\alpha)}\right)$ when $t$ is small, and it behaves asymptotically to $\frac{t^{-\alpha}}{\Gamma(1-\alpha)}$ for large $t$. The stretched exponential function is the same as the Kohlrausch-Williams-Watts (KWW) function [22,23]. a frequently used time correlation function for segmental motion in macromolecular systems [4,5]. There is a significant distinction in the relaxation rate expression between normal and anomalous relaxation [7,13,14,16]. The anomalous relaxation rate expression obtained from the MLF-based time correlation function has been used to successfully fit the experimental data [7,16]. Compared to the traditional empirical KWW function, the MLF-based relaxation time expression uses less fitting parameters [7,16]. Some readers may not be familiar with fractional diffusion. They can skip the fractional diffusion content, which will not affect their understanding of normal phase diffusion results in this paper. However, fractional diffusion is a convenient tool to analyze nonlinear phenomena that exist broadly in biological and polymer systems.

From the normal and fractional phase random walks in this work, the coupling between phase and time can significantly change the spectral density term, which appears in the expression of the obtained phase diffusion coefficient. In traditional theory, the spectral density is obtained from the Fourier transform of the time correlation function of the Hamiltonian interactions [1,2], while the correspondingly apparent spectral density can be directly extracted from the diffusion coefficient in the rotating frame reference for the phase diffusion [7]. The rotating frame evolves at the same frequency as the spin operator evolves [1,2,7]. The NMR relaxation rate is proportional to the spectrum density [1,2,7]. This paper obtains the effective phase diffusion coefficient for both coupled and uncoupled random walk or phase diffusion, in a static frame or a rotating frame. It is found that the spectral density appearing in the obtained diffusion coefficient is significantly affected by the phase-time coupling; the angular frequency is modified from $\omega$ to an apparent angular frequency $\eta\omega$ in the spectrum density expression; this change could significantly affect the analysis result obtained from NMR relaxation experiments. The modified relaxation expression for dipolar coupling is used as an example, which successfully fits the experimental data for $^{13}$C and $^{1}$H coupling NMR relaxation time previously reported in Ref. [6]. The results here give additional insights into NMR relaxation, which could improve the analysis of NMR and magnetic resonance imaging (MRI) experiments in various systems, such as polymer and biological systems.

## II. THEORY
### A. Phase random walk under random field

The random molecular motion alerts the relative molecular orientations, modulating many fundamental Hamiltonians of spin systems, which can be viewed as a random field influencing the evolution of pertinent spin moments [1,2,7]. A simplified random field $H_1(t)$ can be used to show how a fluctuating field affects spin relaxation. $H_1(t)$ can be given by [2,7]

$$H_1(t) = \sum_{q=x,y,z} H_q(t) I_q, \qquad (1)$$

where $I_q$ is the component of the angular momentum, and the amplitude of $H_q(t)$ of the random field can be described by

$$|H_q(t)| \propto \gamma \hbar h_q, \qquad (2)$$

where $\gamma$ is the gyromagnetic ratio, $\hbar$ is the reduced Planck constant, and $h_q$ is the magnetic field intensity. The amplitude of $H_q(t)$ is proportional to $\gamma\hbar h_q$, but changes its direction randomly after each interval $t_i$. Affected by the random field, the spin system undergoes a random phase walk. If all the intervals have the same fixed length, this random walk is a simple diffusion case. During the interval $t_i$, the phase change is [2,7,8]

$$|\Delta\phi_i| = \omega_0 t_i, \qquad (3a)$$

$$\omega_0 \propto \gamma\hbar h_q. \qquad (3b)$$

$\Delta\phi_i$ can be positive or negative depending on the sign of $H_q(t)$ [2,7]. Under the magnetic field, the term of the Hamiltonian interaction inducing the random field precess at a relative frequency $\omega$ to the magnetization components that are observed in the NMR relaxation experiments. Because the observed magnetizationare vectors from the ensemble of spin moments, for each spin, only one component of the spin moment contributes to the observable of the spin ensemble. Therefore, the rotating Hamiltonian interaction leads to an accumulated net phase change of the observable during the interval $t_i$, which could be approximately described as

$$|\Delta\phi_i| \approx \int_0^{t_i} \omega_0 \cos(\omega t)\, dt = \omega_0 \frac{\sin(\omega t)}{\omega}, \qquad (4)$$

where $\omega_0 \cos(\omega t)$ is the projection of the rotating $H_1(t)$ to its starting position of each random jump. The projection of the Hamiltonian interaction by the relative rotating frequency could be understood by a quantum mechanics description (see Appendix A) as well as a classical description. Here, a simple classical instance could help us understand the projection: Assuming there is a magnetic bar $I$ in the $z$ direction, a random field applied along the $x$-axis leads the magnetic bar $I$ to rotate away from the $z$-axis toward the $y$-axis, while if the random magnetic field is applied along the $-x$ axis, it leads the magnetic bar $I$ to rotates away from $z$-axis toward the $-y$ axis. Further, if the direction of the random field is rotating inside the $xy$ plane, the absolute value of the angle or the phase $\Delta\phi_i$ that it moves away from the $z$ direction during the jump time could be approximately proportional to $|\Delta\phi_i| \approx \int_0^{t_i} \omega_0 \cos(\omega t)\, dt$. More delicate approximations could be obtained with further research effort.

The relative starting positions of the Hamiltonian interactions affect the evolution of the observables in NMR relaxation experiments. The spatial average of the starting positions from all spins has been considered and included in the NMR time expressions from the traditional NMR relaxation theory [1], and the same spatial average strategies will be adopted to obtain relaxation rate expressions and will not be focused on in this paper.

When $\omega = 0$, $\omega_0 \frac{\sin(\omega t_i)}{\omega} = \omega_0 t_i$, which can be viewed as a specific case with no projection. In contrast, the Fourier transform of the time correlation function in traditional second-order perturbation theory [1] can be viewed as a specific kind of projection, equivalent to $\frac{\langle \omega_0^2 \tau_i^2 \rangle}{2\langle \tau_i \rangle} \frac{\int_0^\infty \cos(\omega t') G(t) dt}{\int_0^\infty G(t) dt}$ [1,2,7], but it projects the average square of the phase shifts in the ensemble spin system, $\langle \omega_0^2 \tau_i^2 \rangle$, which could overcount the phase shift contribution from the random jump with a long waiting interval. For a rotating Hamiltonian, only the last incomplete phase shift cycle gives the net phase shift contribution, which implies jump with a long waiting time does not necessarily offer considerable net phase jump length in the rotating frame.

It is important to distinguish between the random field and the radio frequency (r.f.) field. The r.f. field is continuously applied to the ensemble spins, while the random field exerts on an individual spin. As the random field acts on the individual spin, only a particular component in the evolution of the individual spin needs to be considered for its contribution to the corresponding observable of the ensemble; other components or directions do not contribute to the observable vector of the spins' ensemble. An instantaneous projection is thus employed for each spin affected by the random field.

As the general phase diffusion has been derived in Ref. [7]. We shall focus on the effect of phase-time coupling on the diffusion coefficient, which is equivalent to the relaxation rate.

### B. Phase-time coupled diffusion

The coupled phase random walk has a joint probability function $\psi(\phi, t)$ defined by [9,11,12]

$$\psi(\phi, t) = \varphi(t)\Phi(\phi|t), \tag{5}$$

where $\varphi(t)$ is the waiting time function, $\Phi(\phi|t)$ is the conditional probability that a phase jump length $\phi$ requires time $t$. In the static frame, the conditional probability is [7,9,10,15]

$$\Phi_{static}(\phi|t) = \frac{1}{2}\delta(|\phi| - \omega_0 t). \tag{6}$$

where $\omega_0 t$ is the absolute value of spin phase change. Besides the rotating frame, the static frame is needed to be investigated because some Hamiltonian interactions' components, such as the $I_z$ is not affected by the rotating frame. In the rotating frame reference, based on Eqs. (4) and (5), the joint conditional probability is

$$\Phi_{rotate}(\phi|t) = \frac{1}{2}\delta(|\phi| - \omega_0 \int_0^t dt'\, \cos(\omega t')) = \frac{1}{2}\delta(|\phi| - \omega_0 \frac{\sin(\omega t)}{\omega}), \tag{7}$$

In Fourier-Laplace representation, the probability density function $P(k, s)$ of a coupled random walk has been derived in Ref. [10,12] as

$$P(k,s) = \frac{\Psi_{jn}(k,s)}{1-\psi(k,s)}, \tag{8}$$

where $\Psi_{jn}(k,s)$ is the Fourier-Laplace representation of the PDF of joint probability $\Psi_{jn}(\phi,t)$ for the phase displacement of the last incomplete walk. $\Psi_{jn}(\phi,t)$ is defined by [9,11,12]

$$\Psi_{jn}(\phi,t) = \Phi(\phi|t)\Psi_{sv}(t), \tag{9a}$$

$$\Psi_{sv}(t) = \int_t^\infty \varphi(t')dt', \tag{9b}$$

where $\Psi_{sv}(t)$ is the survival probability of random walk [10-12], whose Laplace representation is [10-11]

$$\Psi_{sv}(s) = \frac{1-\varphi(s)}{s}. \tag{10}$$

In the static frame, the Laplace-Fourier representation of $\Psi_{jn}(\phi,t)$ for coupled diffusion in the static frame can be calculated from Eqs. (6), (9), and (10) as [9,11]

$$\Psi_{jn,static}(k,s) = \iint e^{ik\phi-st}\Psi_{jn,static}(\phi,t)d\phi dt$$

$$= \frac{1}{2}\iint e^{ik\phi-st}[\delta(\phi+\omega_0 t) + \delta(\phi-\omega_0 t)]\Psi_{sv}(t)d\phi dt$$

$$= \frac{1}{2}[\Psi_{sv}(s+ik\omega_0) + \Psi_{sv}(s-ik\omega_0)]. \tag{11a}$$

Similarly, the Laplace-Fourier representation of the joint probability function $\psi(\phi,t)$ in the static frame is calculated based on Eqs. (5) and (6) [9,11]:

$$\psi_{static}(k,s) = \iint e^{ik\phi-st}\psi(\phi,t)d\phi dt = \frac{1}{2}[\varphi(s+ik\omega_0) + \varphi(s-ik\omega_0)]. \tag{11b}$$

Eqs. (11a) and (11b) can be substituted into Eq. (8) to give

$$P_{c,static}(k,s) = \frac{\frac{1}{2}[\Psi_{sv}(s+ik\omega_0)+\Psi_{sv}(s-ik\omega_0)]}{1-\frac{1}{2}[\varphi(s+ik\omega_0)+\varphi(s-ik\omega_0)]}, \tag{11c}$$

which has been given in Ref. [12].

While in the rotating frame, $\Psi_{jn,rotate}(k,s)$, the Laplace-Fourier representation of the joint survival probability $\Psi_{jn}(\phi,t)$ for coupled diffusion can be calculated based on Eqs. (6), (7), and (10) as

$$\Psi_{jn,rotate}(k,s) = \iint e^{ik\phi-st}\Psi_{jn}(\phi,t)d\phi dt$$

$$= \frac{1}{2}\iint e^{ik\phi-st}\left[\delta\left(\phi+\omega_0\frac{\sin(\omega t)}{\omega}\right) + \delta\left(\phi-\omega_0\frac{\sin(\omega t)}{\omega}\right)\right]\Psi_{sv}(t)d\phi dt$$

$$= \frac{1}{2}\int\left[e^{ik\omega_0\frac{\sin(\omega t)}{\omega}-st} + e^{-ik\omega_0\frac{\sin(\omega t)}{\omega}-st}\right]\Psi_{sv}(t)dt = \int\left[\cos\left(k\omega_0\frac{\sin(\omega t)}{\omega}\right)e^{-st}\right]\Psi_{sv}(t)dt$$

$$\xrightarrow{\cos\left(k\omega_0\frac{\sin(\omega t)}{\omega}\right)\approx 1-\frac{1}{2}\left(\frac{k\omega_0}{\omega}\right)^2\frac{1}{2}(1-\cos(2\omega t))}\left[1-\frac{1}{4}\left(\frac{k\omega_0}{\omega}\right)^2\right]\Psi_{sv}(s) + \frac{1}{4}\left(\frac{k\omega_0}{\omega}\right)^2\frac{1}{2}[\Psi_{sv}(s+i2\omega) + \Psi_{sv}(s-i2\omega)], \tag{12}$$

where $\omega$ is the angular frequency of the rotating frame reference resulting from the external magnetic field, while $\omega_0$ is the angular frequency arising from the Hamiltonian interaction. The approximation in Eq. (12) is based on that $\frac{\omega_0}{\omega}$ is small in NMR relaxation, considering that the frequency of most NMR spectroscopy is hundreds of MHz, while the value of $\omega_0$ is tens of kHz or even smaller. Similarly, based on Eqs. (5) and (7) [9,11], $\psi_{rotate}(k,s)$, the Laplace-Fourier representation of the joint probability $\psi(\phi,t)$ can be calculated

as

$$\psi_{rotate}(k,s) = \iint e^{ik\phi - st}\psi(\phi,t)d\phi dt = \frac{1}{2}\iint e^{ik\phi - st}\left[\delta\left(\phi - \omega_0 \frac{\sin(\omega t)}{\omega}\right) + \delta\left(\phi - \omega_0 \frac{\sin(\omega t)}{\omega}\right)\right]\varphi(t)d\phi dt =$$
$$\frac{1}{2}\int\left[e^{ik\omega_0\frac{\sin(\omega t)}{\omega} - st} + e^{-ik\omega_0\frac{\sin(\omega t)}{\omega} - st}\right]\varphi(t)dt = \int \cos\left(k\omega_0\frac{\sin(\omega t)}{\omega}\right)e^{-st}\varphi(t)dt$$
$$\xrightarrow{\cos\left(k\omega_0\frac{\sin(\omega t)}{\omega}\right)\approx 1-\frac{1}{2}\left(\frac{k\omega_0}{\omega}\right)^2\frac{1}{2}(1-\cos(2\omega t))}\left[1 - \frac{1}{4}\left(\frac{k\omega_0}{\omega}\right)^2\right]\varphi(s) + \frac{1}{4}\left(\frac{k\omega_0}{\omega}\right)^2\frac{1}{2}[\varphi(s+i2\omega) + \varphi(s-i2\omega)].$$
(13)

Eqs. (12) and (13) can be substituted into Eq. (8) to give

$$P_{c,rotate}(k,s) = \frac{\left[1-\frac{1}{4}\left(\frac{k\omega_0}{\omega}\right)^2\right]\Psi_{sv}(s) + \frac{1}{4}\left(\frac{k\omega_0}{\omega}\right)^2\frac{1}{2}[\Psi_{sv}(s+i2\omega) + \Psi_{sv}(s-i2\omega)]}{1 - \left\{\left[1-\frac{1}{4}\left(\frac{k\omega_0}{\omega}\right)^2\right]\varphi(s) + \frac{1}{4}\left(\frac{k\omega_0}{\omega}\right)^2\frac{1}{2}[\varphi(s+i2\omega) + \varphi(s-i2\omega)]\right\}}.$$
(14)

Eq. (14) results from the vector's projection into the rotating frame.

It is worth noting: Compared to Ref. [10], $k = 1$ needs to be considered in the calculation throughout this paper because the NMR magnetization $M(t) = \int_{-\infty}^{\infty} d\phi\, e^{i\phi} P(\phi,t)$ where $k = 1$ is needed for the average in phase space.

### 1. Coupled normal diffusion

#### (i). Static frame

If the random motion's time correlation function is $G(t) = \exp\left(-\frac{t}{\tau}\right)$ [1,2,7], a monoexponential function, the waiting time distribution may be obtained as [24]

$$\varphi(t) = -\frac{dG(t)}{dt} = \frac{1}{\tau}\exp\left(-\frac{t}{\tau}\right),$$
(15)

whose Laplace representation is [11]

$$\varphi(s) = \frac{1}{\tau s + 1},$$
(16)

Eqs. (6) and (16) can be substituted into Eqs. (10) and (11a-c) to give the Laplace-Fourier domain PDF $P_{c,n,static}(k,s)$ [9]:

$$P_{c,n,static}(k,s) = \frac{\Psi_{jn,static}(k,s)}{1-\psi_{static}(k,s)} = \frac{\frac{\tau(1+\tau s)}{(\tau s+1)^2 + k^2\omega_0^2\tau^2}}{1 - \frac{1+\tau s}{(\tau s+1)^2 + k^2\omega_0^2\tau^2}}.$$
(17)

When $k = 1$, [7]

$$P_{c,n,static}(k,s)\Big|_{k=1} = \frac{\frac{\tau(1+\tau s)}{(\tau s+1)^2 + \omega_0^2\tau^2}}{1 - \frac{1+\tau s}{(\tau s+1)^2 + \omega_0^2\tau^2}} \approx \frac{\tau(1+\tau s)}{\omega_0^2\tau^2 + \tau s} = \frac{\frac{\tau}{\omega_0^2\tau^2}}{\frac{\tau s(1-\omega_0^2\tau^2)}{\omega_0^2\tau^2} + 1} = \frac{1}{(1-\omega_0^2\tau^2)}\frac{1}{s+\frac{\omega_0^2\tau}{(1-\omega_0^2\tau^2)}} \approx \frac{1}{s+\frac{\omega_0^2\tau}{(1-\omega_0^2\tau^2)}},$$
(18)

which is the NMR signal because the net magnetization, $M(t) = \int_{-\infty}^{\infty} d\phi\, e^{i\phi} P(\phi,t) = P_{c,n,static}(k,s)\Big|_{k=1}$ in the Laplace domain. In Eq. (18), $\omega_0^2\tau^2 \ll 1$ usually holds, so $\frac{1}{(1-\omega_0^2\tau^2)} \approx 1$. From Eq. (18), it is evident that the phase diffusion coefficient $D_{\phi,c,n,static}$ for coupled normal diffusion in the static frame is

$$D_{\phi,c,n,static} = \frac{\omega_0^2\tau}{(1-\omega_0^2\tau^2)} \approx \omega_0^2\tau,$$
(19)

which replicates Eq. (33), the result of uncoupled normal diffusion presented in Section 2.3.1.

### *(ii). Rotating frame*

Based on Eqs. (10), (12-14), and (16), the Laplace-Fourier domain PDF $P_{c,n,rotate}(k,s)$ for normal diffusion can be given by

$$P_{c,n,rotate}(k,s) = \frac{\left[1 - \frac{1}{4}\left(\frac{k\omega_0}{\omega}\right)^2\right]\tau + \frac{1}{4}\left(\frac{k\omega_0}{\omega}\right)^2 \frac{1}{2}\left[\frac{1 - \frac{1}{\tau(s+i2\omega)+1}}{s+i2\omega} + \frac{1 - \frac{1}{\tau(s-i2\omega)+1}}{s-i2\omega}\right]}{1 - \left\{\left[1 - \frac{1}{4}\left(\frac{k\omega_0}{\omega}\right)^2\right]\frac{1}{\tau s+1} + \frac{1}{4}\left(\frac{k\omega_0}{\omega}\right)^2 \frac{1}{2}\left[\frac{1}{\tau(s-i2\omega)+1} + \frac{1}{\tau(s+i2\omega)+1}\right]\right\}}$$

$$= \frac{\left[1 - \frac{1}{4}\left(\frac{k\omega_0}{\omega}\right)^2\right]\tau + \frac{1}{4}\left(\frac{k\omega_0}{\omega}\right)^2 \frac{\tau(1+\tau s)}{(\tau s+1)^2 + 4\omega^2\tau^2}}{1 - \left\{\left[1 - \frac{1}{4}\left(\frac{k\omega_0}{\omega}\right)^2\right]\frac{1}{\tau s+1} + \frac{1}{4}\left(\frac{k\omega_0}{\omega}\right)^2 \frac{1+\tau s}{(\tau s+1)^2 + 4\omega^2\tau^2}\right\}}$$

$$\approx \frac{(1+4\omega^2\tau^2 - k^2\omega_0^2\tau^2)}{k^2\omega_0^2\tau} \frac{1}{1 + \left[\frac{\tau(1+4\omega^2\tau^2 - k^2\omega_0^2\tau^2)}{k^2\omega_0^2\tau^2} - \frac{\tau}{1+4\omega^2\tau^2 - k^2\omega_0^2\tau^2}\right]s}$$

$$\approx \frac{(1+4\omega^2\tau^2 - k^2\omega_0^2\tau^2)}{k^2\omega_0^2\tau} \frac{1}{1+\frac{\tau(1+4\omega^2\tau^2 - k^2\omega_0^2\tau^2)}{k^2\omega_0^2\tau^2}s} \approx \frac{1}{\frac{k^2\omega_0^2\tau}{(1+4\omega^2\tau^2 - k^2\omega_0^2\tau^2)}+s}. \tag{20}$$

NMR signal is the average result in the phase space, $\int_{-\infty}^{\infty} d\phi\, e^{i\phi}P(\phi,t) = P(k,t)$ for $k=1$ [7]. Therefore, in the Fourier-Laplace representation, the NMR signal corresponds to $P_{rotate,c,n}(k,s)$ with $k=1$, which is

$$P_{c,n,rotate}(1,s) = \frac{1}{\frac{\omega_0^2\tau}{(1+4\omega^2\tau^2 - \omega_0^2\tau^2)}+s}, \tag{21}$$

and the phase diffusion coefficient $D_{\phi,c,n,rotate}$ for coupled normal diffusion in rotating frame reference is

$$D_{\phi,c,n,rotate} = \frac{\omega_0^2\tau}{1+(2\omega)^2\tau^2 - \omega_0^2\tau^2} \approx \frac{\omega_0^2\tau}{1+(2\omega)^2\tau^2}, \tag{22}$$

where the approximation is based on that $\omega_0^2\tau^2$ can be neglected in the denominator, as $\omega_0 \ll \omega$ and $\omega_0^2\tau^2 \ll 1$ are often satisfied. Instead of angular frequency $\omega$, an apparent angular frequency $2\omega$ appears strikingly in Eq. (22), which results from the phase-time coupling. The phase due to the coupling in this paper is $\omega_0 \int_0^{\tau_i} dt'\, cos(\omega t') = \omega_0 \frac{sin(\omega\tau_i)}{\omega}$ which reduces to $\omega_0 t$ when $\omega$ approaches 0, and the diffusion coefficient is $\langle\left(\omega_0\frac{sin(\omega\tau_i)}{\omega}\right)^2\rangle/(2\langle\tau_i\rangle)$. While in the traditional method, the relaxation rate can be viewed as obtaining equivalently by $\frac{\langle\omega_0^2\tau_i^2\rangle}{2\langle\tau_i\rangle}\frac{\int_0^\infty cos(\omega t')G(t)dt}{\int_0^\infty G(t)dt}$, which may not be a reasonable; when the time $\tau_i$ is infinite, the phase shift from $\omega_0\tau_i$ is infinite; however, the net phase shift is only obtained from the last incomplete rotating cycle, which is $= \omega_0 \frac{sin(\omega\tau_i)}{\omega}$. When $\omega$ is large, the net phase shift is small.

### *2. Coupled fractional diffusion*
### *(i) Static frame*

The time-fractional phase diffusion with an MLF-based waiting time distribution is investigated here. The Laplace representation of MLF-based waiting time distribution [9-11] is

$$\varphi_f(s) = \frac{1}{s^\alpha\tau^\alpha + 1}. \tag{23}$$

Eqs. (11b) and (23) can be combined to calculate the joint probability function's Laplace-Fourier representation in the static frame

$$\psi_{static}(1,s) = \frac{c}{1+s\tau'}, k = 1, \tag{24a}$$

where $k = 1$ because the NMR signal is the average over the distribution of phase space [7-9], and

$$c = \frac{\omega_0^\alpha \tau^\alpha \left(\cos\frac{\pi}{2}\alpha + \frac{1}{\omega_0^\alpha \tau^\alpha}\right)}{1+\omega_0^{2\alpha}\tau^{2\alpha}+2\omega_0^\alpha \tau^\alpha \cos\frac{\pi}{2}\alpha}, \tag{24b}$$

and

$$\tau' = \frac{\alpha \omega_0^{\alpha-1} \tau^\alpha \sin\frac{\pi}{2}\alpha \frac{1-\omega_0^{2\alpha}\tau^{2\alpha}}{\omega_0^\alpha \tau^\alpha \cos\frac{\pi}{2}\alpha + 1}}{1+\omega_0^{2\alpha}\tau^{2\alpha}+2\omega_0^\alpha \tau^\alpha \cos\frac{\pi}{2}\alpha}. \tag{24c}$$

Additionally, Eqs. (11a) and (23) can be combined to calculate the Laplace-Fourier representation of $\Psi_{jn}(\phi, t)$ in the static frame [9]

$$\Psi_{jn,static}(1,s) = \frac{1}{2}\left[\frac{1-\frac{1}{(s+i\omega_0)^\alpha \tau^\alpha+1}}{s+i\omega_0} + \frac{1-\frac{1}{(s-i\omega_0)^\alpha \tau^\alpha+1}}{s-i\omega_0}\right] = \frac{c_1}{1+s\tau'_1}, \tag{25a}$$

where

$$c_1 = \frac{\tau^\alpha \omega_0^{\alpha-1} \sin\frac{\pi}{2}\alpha}{\omega_0^{2\alpha}\tau^{2\alpha}+2\omega_0^\alpha \tau^\alpha \cos\frac{\pi}{2}\alpha+1}, \tag{25b}$$

$$\tau'_1 = \frac{2\alpha \omega_0^{\alpha-1}\tau^\alpha \sin\frac{\pi}{2}\alpha}{\left(\omega_0^{2\alpha}\tau^{2\alpha}+2\omega_0^\alpha \tau^\alpha \cos\frac{\pi}{2}\alpha+1\right)} - \frac{\omega_0^\alpha \tau^\alpha - (\alpha-1)\cos\left(\frac{\pi}{2}\alpha\right)}{\omega_0 \sin\frac{\pi}{2}\alpha}. \tag{25c}$$

Eqs. (24) and (25) can be substituted into Eq. (8) to give

$$P_{c,f,static}(k,s)\big|_{k=1} = \frac{\Psi_{jn,static}(1,s)}{1-\psi_{static}(1,s)} = \frac{\frac{c_1}{1+s\tau'_1}}{1-\frac{c}{1+s\tau'}} \approx \frac{c_1}{(1+s\tau'_1)[1-c(1-s\tau')]} = \frac{c_1}{(1-c)\tau'_1 + c\tau'} \cdot \frac{1}{s + \frac{1}{\tau'_1 + \frac{c}{1-c}\tau'}}. \tag{26}$$

From Eq. (26), it is evident that the phase diffusion coefficient $D_{\phi,c,f,static}$ for coupled fractional diffusion in the static frame is

$$D_{\phi,c,f,static} = \frac{1}{\tau'_1 + \frac{c}{1-c}\tau'}, \tag{27}$$

which reduces to Eq. (19) for coupled normal diffusion when $\alpha = 1$.

### *(ii) Rotating frame*

Based on Eqs. (10), (14), and (23), the PDF for coupled normal diffusion in the rotating frame can be obtained as

$$P_{c,f,rotate}(k,s) = \frac{\left[1-\frac{1}{4}\left(\frac{k\omega_0}{\omega}\right)^2\right]\tau^\alpha s^{\alpha-1} + \frac{1}{4}\left(\frac{k\omega_0}{\omega}\right)^2 \frac{c_1}{1+s\tau'_1}}{1-\left\{\left[1-\frac{1}{4}\left(\frac{k\omega_0}{\omega}\right)^2\right]\frac{1}{s^\alpha \tau^\alpha+1} + \frac{1}{4}\left(\frac{k\omega_0}{\omega}\right)^2 \frac{c}{1+s\tau'}\right\}}, \tag{28}$$

where the constants $c, \tau', c_1, \tau'_1$ are defined by Eqs. (24b), (24c), (25b), and (25c), respectively, but the $\omega_0$ in these expressions needed to be replaced with $2\omega$ for Eq. (28). Further effort is still needed to obtain an apparent phase diffusion coefficient from Eq. (28).

### C. Phase and time uncoupled diffusion

For uncoupled diffusion, the Fourier-Laplace representation of the probability density function is [11,12]

$$P(k,s) = \frac{1-\varphi(s)}{s} \frac{1}{1-\Phi(k)\varphi(s)}, \tag{29}$$

where $\varphi(s)$ is the Laplace representation of the waiting time distribution and $\Phi(k)$ is the distribution of phase jump length.

### 1. Uncoupled normal diffusion

*(i) Static frame*

In the static frame, the phase jump length distribution $\Phi(\phi)$ is assumed to be

$$\Phi(\phi) = \frac{1}{2}\frac{1}{\phi_0}\exp\left(-\frac{|\phi|}{\phi_0}\right), \quad \phi_0=\omega_0\tau. \tag{30}$$

Here, we assume the phase distribution is the same as the time distribution $\frac{1}{\tau}\exp\left(-\frac{t}{\tau}\right)$ based on Eq. (3a), which arises from the natural coupling between phase precession and the interaction time. The Fourier transform of $\Phi(\phi)$ gives

$$\Phi(k) = \psi(k,s) = \int_{-\infty}^{\infty} e^{ik\phi}\Phi(\phi)d\phi = \int_0^\infty e^{ik\phi}\frac{1}{\phi_0}\exp\left(-\frac{\phi}{\phi_0}\right)d\phi = \frac{1}{\phi_0}\frac{\phi_0}{\phi_0^2 k^2+1} = \frac{1}{\phi_0^2 k^2+1}. \tag{31}$$

The waiting time distribution for uncoupled normal diffusion is still the monoexponential function, and its Laplace representation of waiting time distribution is given by Eq. (16). Eqs. (16) and (31) can be substituted into Eq. (29) to obtain

$$P(k,s) = \frac{1-\varphi(s)}{s}\frac{1}{1-\Phi(k)\varphi(s)} = \frac{\tau}{1-\frac{1}{\phi_0^2 k^2+1}\cdot\frac{1}{\tau s+1}} \approx \frac{\tau}{1-(1-\phi_0^2 k^2)(1-s\tau)} \approx \frac{\tau}{\phi_0^2 k^2+s\tau} = \frac{1}{\frac{\phi_0^2 k^2}{\tau}+s}, \tag{32}$$

where the approximations are based on that $\phi_0^2 k^2 = (\omega_0 t)^2 k^2$ and $s\tau$ are small. Eq. (32) implies that the phase diffusion coefficient for uncoupled normal diffusion in the static frame is

$$D_{\phi,uc,n,static} = \frac{\phi_0^2}{\tau} = \omega_0^2\tau. \tag{33}$$

The same phase diffusion coefficient can be obtained alternatively by [7-9]

$$D_{\phi,uc,n,static} = \frac{\langle\phi_0^2\rangle}{2\langle\tau_{jump}\rangle} = \frac{\int_0^\infty(\omega_0 t)^2\varphi(t)dt}{2\int_0^\infty t\varphi(t)dt} = \frac{\int_0^\infty(\omega_0 t)^2\frac{1}{\tau}\exp\left(-\frac{t}{\tau}\right)dt}{2\int_0^\infty\frac{t}{\tau}\exp\left(-\frac{t}{\tau}\right)dt} = \frac{2\omega_0^2\tau^2}{2\tau} = \omega_0^2\tau. \tag{34}$$

*(ii) Rotating frame*

In a rotating frame, the effective phase jump length during an interval $\tau$ is $\omega_0\int_0^\tau dt'\cos(\omega t')$, which can be combined with the monoexponential time distribution Eq. (3a) to obtain the phase diffusion coefficient for the uncoupled normal diffusion in the rotating frame [7-9]

$$D_{\phi,uc,n,rotate} = \frac{\langle\phi_0^2\rangle}{2\langle\tau_{jump}\rangle} = \frac{\int_0^\infty\left(\omega_0\int_0^t dt'\cos(\omega t')\right)^2\varphi(t)dt}{2\int_0^\infty t\varphi(t)dt} = \frac{\int_0^\infty\left(\omega_0\frac{\sin(\omega t)}{\omega}\right)^2\frac{1}{\tau}\exp\left(-\frac{t}{\tau}\right)dt}{2\int_0^\infty\frac{t}{\tau}\exp\left(-\frac{t}{\tau}\right)dt} = \frac{\omega_0^2\tau}{1+(2\omega)^2\tau^2}. \tag{35}$$

Eq. (35) agrees with Eq. (22) for the coupled phase diffusion because $\omega_0^2\tau^2$ is negligible compared to $4\omega^2\tau^2$.

### 2. Uncoupled fractional diffusion

The fractional diffusion could have a waiting time distribution $\varphi_f(t) = -\frac{d}{dt}E_\alpha\left(-\left(\frac{t}{\tau}\right)^\alpha\right)$ [24].

*(i) Static frame*

The phase diffusion coefficient in the static frame can be obtained [7-9]

$$D_{\phi,uc,f,static} = \frac{\langle\phi_0^2\rangle}{2\langle\tau_{jump}\rangle} = \frac{\int_0^\infty(\omega_0 t)^2\varphi(t)dt}{2\Gamma(1+\alpha)\tau^\alpha} = \frac{\int_0^\infty(\omega_0 t)^2\left[-\frac{d}{dt}E_\alpha\left(-\left(\frac{t}{\tau}\right)^\alpha\right)\right]dt}{2\Gamma(1+\alpha)\tau^\alpha}, \tag{36}$$

which divergences. This issue is one of the reasons that has urged researchers to develop the coupled random walk theory [11]. However, if we assume that $\langle \phi_0^2 \rangle = 2(\omega_0 \tau)^2$, which is used in the normal diffusion in Eq. (34), t the phase diffusion coefficient will be

$$D_{\phi,uc,f,static} = \frac{\langle \phi_0^2 \rangle}{2\langle \tau_{jump} \rangle} = \frac{2(\omega_0\tau)^2}{2\Gamma(1+\alpha)\tau^\alpha} = \frac{\omega_0^2 \tau^{2-a}}{\Gamma(1+\alpha)}, \qquad (37)$$

which has a bright point that it reduces to Eqs. (33) and (34) for uncoupled normal diffusion when $\alpha = 1$.

*(ii)* *Rotating frame*

Similarly to Eq. (35), for the uncoupled fractional diffusion [10],

$$D_{\phi,uc,f,rotate} = \frac{\langle \phi_0^2 \rangle}{2\Gamma(1+\alpha)\tau^\alpha} = \frac{\int_0^\infty \left(\omega_0 \int_0^t dt' \cos(\omega t')\right)^2 \varphi(t) dt}{2\Gamma(1+\alpha)\tau^\alpha} = \frac{\int_0^\infty \left(\omega_0 \frac{\sin(\omega t)}{\omega}\right)^2 \left[-\frac{d}{dt} E_\alpha\left(-\left(\frac{t}{\tau}\right)^\alpha\right)\right] dt}{2\Gamma(1+\alpha)\tau^\alpha} =$$

$$\frac{\frac{1}{2}\left(\frac{\omega_0}{\omega}\right)^2 \int_0^\infty (1-\cos(2\omega t))\left[-\frac{d}{dt}E_\alpha\left(-\left(\frac{t}{\tau}\right)^\alpha\right)\right] dt}{2\Gamma(1+\alpha)\tau^\alpha} = \frac{1}{2}\left(\frac{\omega_0}{\omega}\right)^2 \cdot \frac{1 - \int_0^\infty \cos(2\omega t)\left[-\frac{d}{dt}E_\alpha\left(-\left(\frac{t}{\tau}\right)^\alpha\right)\right] dt}{2\Gamma(1+\alpha)\tau^\alpha}$$

$$= \frac{1}{2}\left(\frac{\omega_0}{\omega}\right)^2 \cdot \frac{1 + \cos(2\omega t) E_\alpha\left(-\left(\frac{t}{\tau}\right)^\alpha\right)\Big|_0^\infty + 2\omega \int_0^\infty E_\alpha\left(-\left(\frac{t}{\tau}\right)^\alpha\right) \sin(2\omega t) dt}{2\Gamma(1+\alpha)\tau^\alpha} = \frac{\omega_0^2}{\omega} \frac{\int_0^\infty E_\alpha\left(-\left(\frac{t}{\tau}\right)^\alpha\right) \sin(2\omega t) dt}{2\Gamma(1+\alpha)\tau^\alpha},$$

(38a)

which reduces to Eq. (35) when $\alpha = 1$. Eq. (38a) may be approximated as

$$D_{\phi,uc,f,rotate} \approx \frac{\omega_0^2}{\omega} \frac{2\omega\tau^\alpha \int_0^\infty E_\alpha\left(-\left(\frac{t}{\tau}\right)^\alpha\right) \cos(2\omega t) dt}{2\Gamma(1+\alpha)\tau^\alpha} = \omega_0^2 \frac{2 \int_0^\infty E_\alpha\left(-\left(\frac{t}{\tau}\right)^\alpha\right) \cos(2\omega t) dt}{2\Gamma(1+\alpha)} =$$
$$\omega_0^2 \frac{(2\omega)^{\alpha-1} \tau^\alpha \sin(\pi\alpha/2)}{1 + 2(2\omega\tau)^\alpha \cos(\pi\alpha/2) + (2\omega\tau)^{2\alpha}} \frac{1}{\Gamma(1+\alpha)},$$

(38b)

which reduces to Eq. (35) again when $\alpha = 1$. Compared to the result, $\omega_0^2 \frac{\omega^{\alpha-1}\tau^\alpha \sin(\pi\alpha/2)}{1+2(\omega\tau)^\alpha \cos(\pi\alpha/2)+(\omega\tau)^{2\alpha}}$, in Ref. [7], the apparent angular frequency here is two times greater. The approximation in Eq. (38b) is heuristic, which considers that $E_\alpha\left(-\left(\frac{t}{\tau}\right)^\alpha\right)$ is approximately equal to $\exp\left(-\frac{1}{\Gamma(1+\alpha)}\left(\frac{t}{\tau}\right)^\alpha\right)$ when $\left(\frac{t}{\tau}\right)^\alpha$ is small, and $\int_0^\infty \exp\left(-\frac{t}{\tau}\right) \sin(2\omega t) dt = 2\omega t \int_0^\infty \exp\left(-\frac{t}{\tau}\right) \cos(2\omega t) dt$.

### D. NMR relaxation expressions
#### 1. Phase-time coupling constant

The phase diffusion coefficients of both the coupled and uncoupled normal diffusion from the static frame are $\omega_0^2 \tau$, which is exactly the same as the relaxation rate from traditional theories [2]. However, in the rotating frame, when considering the phase-time coupling, the relaxation rate may be significantly different from the traditional results based on the obtained phase diffusion coefficients.

The results from this model agree with the traditional model except that the angular frequency in the effective phase diffusion coefficient is $2\omega$, which is two times that used in the spectral density for the traditional model if the relative frequency $\omega$ is assumed to be the same as the traditional angular frequency. The apparent angular frequency $2\omega$ appears in both the coupled diffusion and uncoupled diffusion. In coupled diffusion, the coupling is evident through the joint probability function, while the phase-space coupling affects the result of uncoupled diffusion through its effect on the phase variance in the uncoupled random walk, as shown in Eqs. (35) and (38).

The phase-time coupling is handled by the instantaneous projection of the Hamiltonian interaction, while, in traditional theory, the coupling is not considered, and its relaxation rate is equivalently proportional to $\frac{\langle\omega_0^2\tau_i^2\rangle}{2\langle\tau_i\rangle}\frac{\int_0^\infty cos(\omega t')G(t)dt}{\int_0^\infty G(t)dt}$, where $\langle\omega_0^2\tau_i^2\rangle$ can be viewed as the phase variance, an average result of different jump time lengths which, however, may not be reasonable, as the contribution to the effective phase length from a long time jump comes only from the last incomplete cycle, $\omega_0\frac{sin(\omega\tau_i)}{\omega}$, that is much smaller than $\omega_0\tau_i$.

If we assume a phase-time coupling constant $\eta$, from Eqs. (22), (35), for both normal coupled and uncoupled phase diffusion, we have

$$D_{\phi,n,rotate} = \omega_0^2 \frac{\tau}{1+[\eta\omega]^2\tau^2}, 0 \leq \eta \leq 2. \tag{39}$$

While for fractional diffusion, from Eq. (38b),

$$D_{\phi,f,rotate} = \omega_0^2 \frac{((1+\eta)\omega)^{\alpha-1}\tau^\alpha \sin(\pi\alpha/2)}{1+2((1+\eta)\omega\tau)^\alpha \cos(\pi\alpha/2)+((1+\eta)\omega\tau)^{2\alpha}} \frac{1}{\Gamma(1+\alpha)}. \tag{40}$$

$\eta$ =2 corresponds to the strongest coupling, while $\eta$ =1 corresponds to the traditional result. The range of $\eta$ should be from 0 to 2. However, here, it is deliberately set from 0 to 2, as the relative frequency $\omega$ may be smaller than the traditional frequency used in the NMR relaxation expressions, although the possibility of $\eta$ <1 may be small. Because the coupling constant is a motional feature of a spin system, it does not depend on the applied external magnetic field of the NMR spectroscopy.

## 2. NMR relaxation expression example

Here, the NMR spin-lattice relaxation due to dipolar coupling is used as an example to show how the coupling constant $\eta$ changes the relaxation rate expression. Ref. [7] shows that the spin-lattice relaxation rate is equivalent to the phase diffusion constant, namely $\frac{1}{T_1} = D_\phi$. The spin-lattice relaxation expression for dipolar coupling between unlike spins, such as $^1H$ and $^{13}C$ coupling, could be obtained from the phase diffusion results in Ref. [7] by modifying the phase diffusion coefficients by including the phase-time coupling constant $\eta$ to give

$$\frac{1}{T_1} = D_\phi = D_\phi^{(0)} + D_\phi^{(1)} + D_\phi^{(2)} = \frac{2}{15r^6}(\frac{\mu_0}{4\pi}\gamma_I\gamma_S)^2\hbar^2 S(S+1)$$
$$\left\{\frac{\tau}{1+[\eta_0(\omega_I-\omega_S)]^2\tau^2} + \frac{3\tau}{1+[\eta_1\omega_I]^2\tau^2} + \frac{6\tau}{1+[\eta_2(\omega_I+\omega_S)]^2\tau^2}\right\},$$

$$D_\phi^{(0)} = \frac{1}{15r^6}(\frac{\mu_0}{4\pi}\gamma_I\gamma_S)^2\hbar^2 S(S+1)\frac{2\tau}{1+[\eta_0(\omega_I-\omega_S)]^2\tau^2},$$
$$D_\phi^{(1)} = \frac{1}{5r^6}(\frac{\mu_0}{4\pi}\gamma_I\gamma_S)^2\hbar^2 S(S+1)\frac{2\tau}{1+[\eta_1\omega_I]^2\tau^2},$$

$$D_\phi^{(2)} = \frac{2}{5r^6}(\frac{\mu_0}{4\pi}\gamma_I\gamma_S)^2\hbar^2 S(S+1)\frac{2\tau}{1+[\eta_2(\omega_I+\omega_S)]^2\tau^2},$$
$$0 \leq \eta_i \leq 2, i = 1,2,3. \tag{41a}$$

where $I$ and $S$ represent the two coupling spins such as $^{13}C$ and $^1H$, which have the spin numbers $I$ and $S$, respectively, $\omega_I$ and $\omega_S$ are the angular frequencies of the two spins, respectively, $\eta_i$ are the coupling constants for $i^{th}$ order quantum coherences, $D_\phi^{(q)}, q = 1,2\ 3$ are the phase diffusion coefficient resulting from the $q^{th}$ order Hamiltonian interaction [7], and $r$ is the spatial distance between the two spins.    Eq. (41a) reduces to the traditional spin-lattice relaxation expression when $\eta_i$ = 1 [1,3,7]. In Eq. (41a), the relative frequency is assumed to be the frequencies $\omega_I - \omega_S, \omega_I$, and $\omega_I + \omega_S$ that are used by traditional theories; further research could provide improved relaxation time expressions with different relative frequencies for these Hamiltonian interaction terms.    The effect of coupling between phase and time increases when $\eta$ increases.   From Eq. (41a), for $^{13}C$ spin-lattice relaxation experiment [1,3,7], the relaxation time obeys

$$\frac{1}{T_1} = n_H D_\phi, \tag{41b}$$

where $n_H$ is the number of the attached Hydrogen nuclei.

### III. RESULTS AND DISCUSSION

The general phase random walk for a spin system in the NMR relaxation process is investigated based on the coupled and uncoupled CTRW theories. This paper focuses on obtaining the effective phase diffusion coefficients in these different situations, as the effective phase diffusion coefficient can be interpreted as the NMR relaxation rate [7]. The results include both normal diffusion and fractional diffusion. Additionally, all the diffusions are considered in the static frame as well as the rotating frames.

From the obtained effective phase diffusion coefficients, the phase-time coupling leads to a two-time difference in angular frequency appearing in the NMR relaxation rate expressions. In traditional NMR theory, the relaxation rate can be obtained by the second-order perturbation theory [1-3] (see Appendix A). The relaxation rate of the traditional result is proportional to the spectral density of the time correlation function; the spectral density is proportional to $\frac{\tau}{1+\omega^2\tau^2}$ for a monoexponential time correlation function while it is proportional to $\frac{\omega^{\alpha-1}\tau^\alpha \sin(\pi\alpha/2)}{1+2(\omega\tau)^\alpha \cos(\pi\alpha/2)+(\omega\tau)^{2\alpha}}$ for an MLF-based time correlation function. Based on the analysis of the effective phase diffusion coefficients for phase random walks, the relaxation rates in both the coupled and uncoupled normal diffusion are proportional to $\frac{\tau}{1+(2\omega)^2\tau^2}$ for the monoexponential correlation, while for uncoupled fractional diffusion, the relaxation rates are proportional to $\frac{\omega_0^2}{\omega}\frac{\int_0^\infty E_\alpha\left(-\left(\frac{t}{\tau}\right)^\alpha\right)\sin(2\omega t)dt}{2\Gamma(1+\alpha)\tau^\alpha}$. The NMR relaxation rate is fast when the relative motion speed is near the neighborhood of the on-resonance motion where $2\omega\tau = 1$, while it becomes slower when the relative motion is off-resonance. When $\alpha = 1$, the fractional spectral density reduces to the normal spectral density. For the coupled fractional random walk, although the expression $P_{c,f,rotate}(k,s)$ is given by Eq. (28), it still needs further effort to obtain the effective diffusion coefficient. The apparent angular frequency obtained from the model presented in this paper is twice that used by traditional theories. This increase in the apparent angular frequency arises from the following: Because the phase jump length during a jump time $\tau_i$ is proportional to $\omega_0 \int_0^{\tau_i} dt' \cos(\omega t') = \omega_0 \frac{\sin(\omega\tau_i)}{\omega}$, the variance of the phase random walk is $\langle \omega_0^2 \frac{\sin^2(\omega\tau_i)}{\omega^2}\rangle = \langle \omega_0^2 \frac{1-\cos(2\omega\tau_i)}{2\omega^2}\rangle$, where $2\omega\tau_i$ rather than $\omega\tau_i$ appears. While in the traditional theory, the relaxation rate is obtained equivalently by $\frac{\langle\omega_0^2\tau_i^2\rangle}{2\langle\tau_i\rangle}\frac{\int_0^\infty \cos(\omega t')G(t)dt}{\int_0^\infty G(t)dt}$ [7] where the phase variance is average first, then its result is combined with the Fourier transform of correlation time; this average strategy could overcount the long time jump's contribution to the phase variance. A phase-time coupling constant $\eta$ could be proposed to include the effect of phase-time coupling in the NMR relaxation rate; $\eta\omega$ is the apparent angular frequency for NMR relaxation, with $0 \leq \eta \leq 2$. When $\eta = 1$, the results reduce to traditional results. The range of $\eta$ is set from 0 to 2 rather than from 1 to 2, because, currently, it is unknown whether or not $0 \leq \eta < 1$ exists. Further research could provide a more accurate relative frequency for each Hamiltonian interaction. When the relative motion is fast, namely $\tau$ is small, $\frac{\tau}{1+(2\omega)^2\tau^2} \approx \tau$ and the coupling effect is negligible. The fast motion can be observed in small molecule liquid state NMR experiments [1].

The spectral density from a monoexponential function based on conventional theory usually cannot fit the NMR relaxation time in amorphous polymer samples, where the modified KWW function and MLF-based relaxation expression can be applied [6, 7]. The modified KWW function can be described as [7]

$$G_{mKWW}(t) = a_{lib}\exp\left(-\frac{t}{\tau_{lib}}\right)+(1-a_{lib})\exp\left[-\left(\frac{t}{\tau}\right)^\alpha\right], \tag{42a}$$

where $\tau_{lib}$ is the time constant of liberational motion, often set as 1 ps. The KWW function $exp\left[-\left(\frac{t}{\tau}\right)^\alpha\right]$ is often expanded by $exp\left[-\left(\frac{t}{\tau}\right)^\alpha\right] = \sum_i \rho_i exp\left(-\frac{t}{\tau_i}\right)$ where $\rho_i$ are the coefficients. The spectral density of $G_{mKWW}(t)$ is

$$J_{mKWW}(\omega) = a_{lib}\frac{\tau_{lib}}{1+(\eta\omega)^2\tau_{lib}^2} + (1-a_{lib})\sum_i \rho_i \frac{\tau_i}{1+(\eta\omega)^2\tau_i^2}. \tag{42b}$$

When $\eta = 1$, Eq. (42b) is the traditional spectral density for the modified KWW function.

The relaxation time expressions (39-42) based on the apparent angular frequency $\eta\omega$ are applied to fit the experimental $^{13}C$ $T_1$ NMR data taken from Ref. [6]. This $^{13}C$ $T_1$ data are for the methylene group of polyisobutylene (PIB) in 70% PIB and 30% head-to-head poly(propylene) (hhPP) sample, which was measured at variable temperatures and two field frequencies, 50.3 MHz and 100.6 MHz. For simplicity, all the $\eta_i, i = 1,2,3$ in Eq. (41a) are set as the same. For convenience, the subindex *i* of $\eta_i$ will be dropped in all the Figures and throughout the rest of the paper. The fitting results are displayed in Figure 1. The $D_\phi$ from Eq. (39) for the coupled and uncoupled normal diffusion and the $D_\phi$ from Eq. (40) based on the MLF Eq. (40) for the coupled fractional diffusion are used in the fitting. Without the coupling effect, namely $\eta = 1$, with the fixed angular frequency *ω*, the monoexponential time correlation function based on the traditional theory cannot successfully interpret this data.

Figure 2 compared the fitting based on modified KWW functions with $\eta = 1.4$ and $\eta = 1$. The fitting curves with $\eta = 1$ are calculated based on the parameters reported in Ref. [7], which corresponds to the traditional theoretical results, while curves with $\eta_i = 1.4$ represent the results based on the phase-time coupling.

In the fitting, the Vogel-Tamman-Fulcher (VTF) temperature dependence [5,6]:

$$\tau = \tau_\infty \times 10^{\frac{B}{T-T_0}}$$

is used to give the temperature-dependent segmental dynamics, where $\tau_\infty$ is a time scale, *B* is the activation energy divided by the Boltzmann constant, $T_0$ is the Vogel temperature, and *T* is the experimental temperature. The fitting parameters are listed in Table 1. The fittings use four parameters for Eq. (39) but five parameters for Eq. (40). In contrast, the traditional mKWW fitting needs six parameters, $\alpha$, $\tau_\infty$, *B*, $T_0$, $a_{lib}$, and $\tau_{lib}$. Interesting, the modified KWW function with $\eta_i = 1.4$ has a $a_{lib}$ value equaling 0, which implies that the $a_{lib}$ parameter for liberational motion may be unnecessary when coupling constant $\eta_i$ is employed in the fitting. No liberational motion is needed for the MLF-based fitting in this paper and the MLF-based fitting reported in Ref. [7,9].

**Table 1:** $^{13}C$ $T_1$ fitting parameters with coupling constant $\eta$, $0 \leq \eta \leq 2$.

| Dynamic Mode | $\alpha$ | $\tau_\infty$ (ps) | B (K) | $T_0$ (K) | $\eta$ |
|---|---|---|---|---|---|
| Coupled and uncoupled normal diffusion, Eqs. (39) $\alpha = 1$ | | 0.15 | 1032 | 70 | 1.97 |
| Coupled fractional diffusion, Eq. (40) | 0.79 | 0.065 | 1000 | 110 | 1.54 |
| mKWW with coupling | 0.61 | 0.01 | 1250 | 100 | 1.4 |
| mKWW parameters taken from Ref. [6] $\tau_{lib} = 0.1$ ps, $a_{lib} = 0.26$ | 0.6 | 0.1 | 775 | 160 | * |

* This traditional fitting is equivalent to having $\eta = 1$.

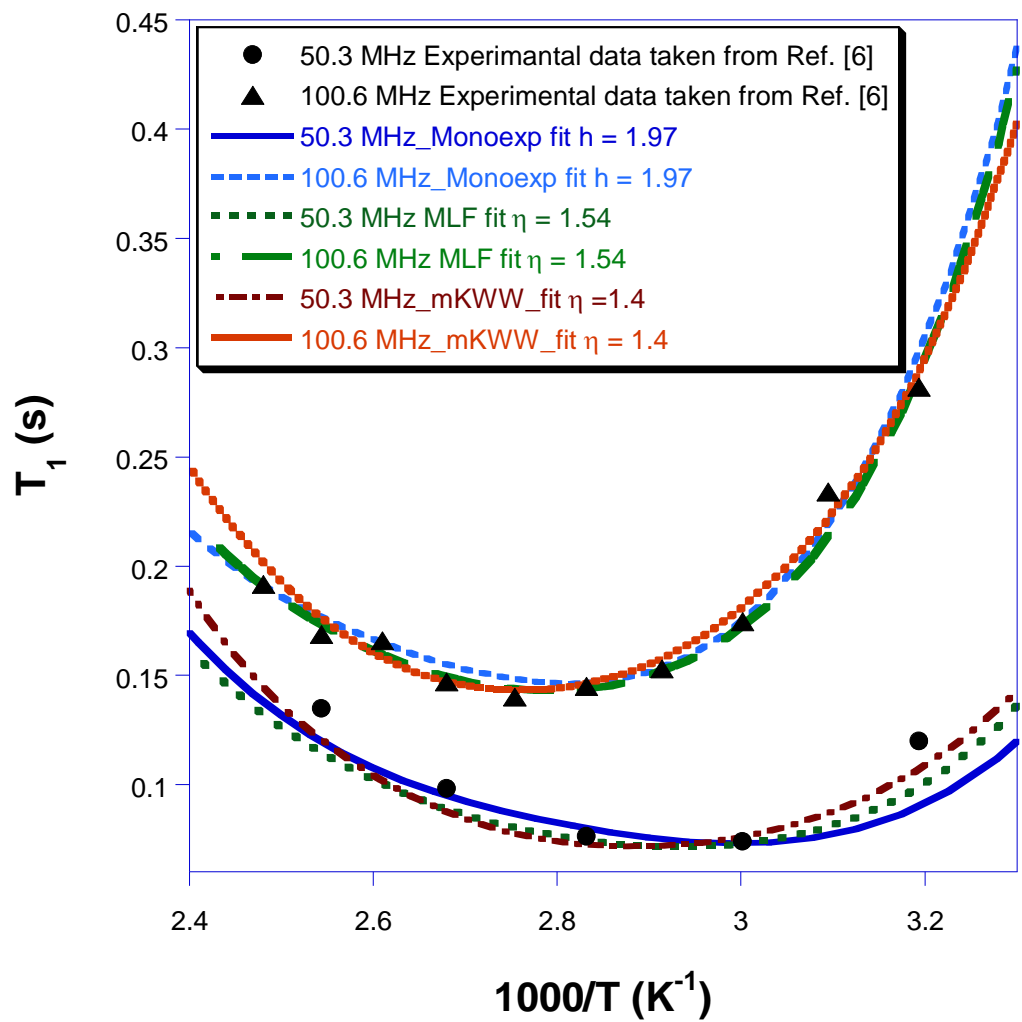

**Figure 1.** Fitting the spin-lattice relaxation time $^{13}$C $T_1$ experimental data by Eqs. (39) and (40). The data are taken from Ref. [6], which is measured at variable temperatures and two field frequencies, 50.3 MHz and 100.6 MHz, for the methylene group of polyisobutylene (PIB) in 70% PIB and 30% head-to-head poly(propylene) (hhPP) sample. Both the monoexponential function, the modified KWW and the MLF-based models can successfully fit the data.

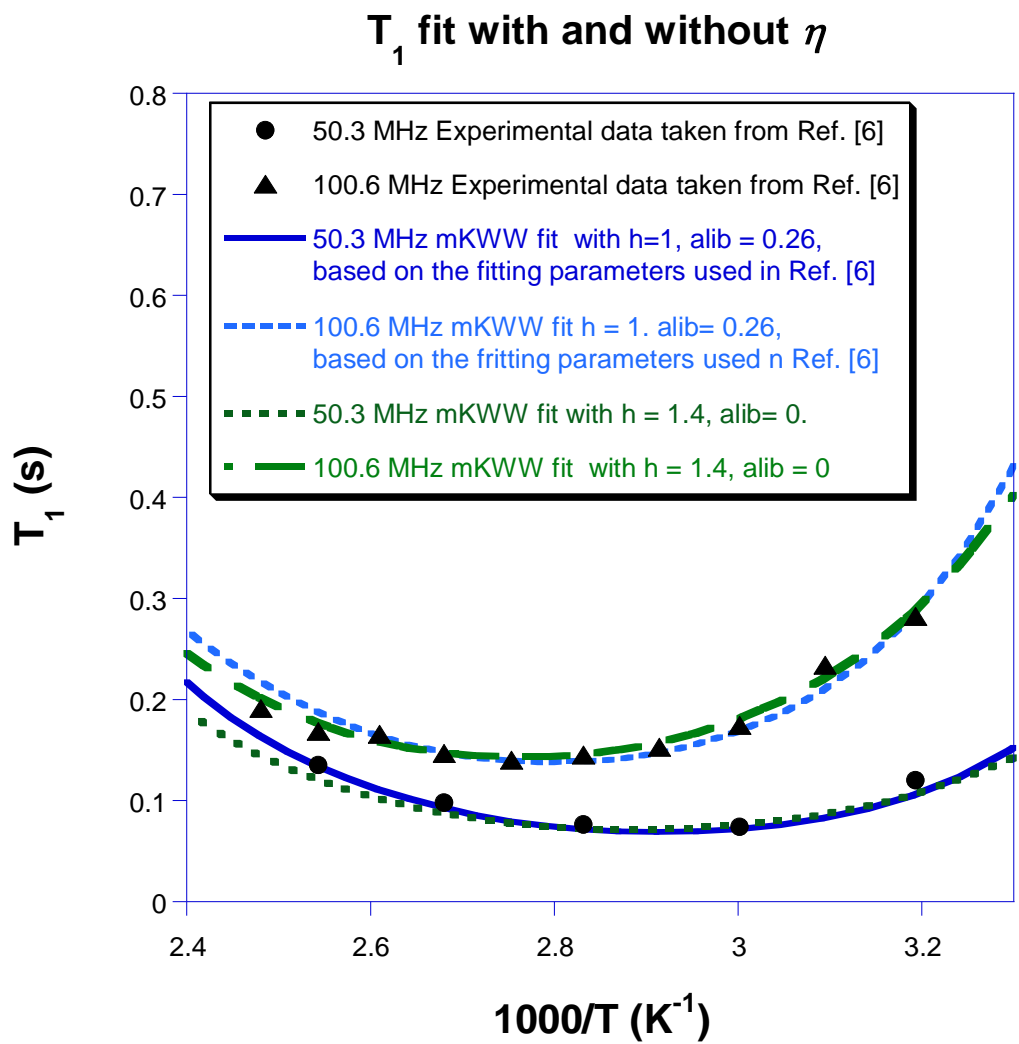

**Figure 2.** Comparison of the fitting based on the modified KWW function with coupling constant and the traditional fitting based on the modified KWW function. $^{13}C$ $T_1$ experimental data are obtained from Ref. [6], which is measured at variable temperatures and two field frequencies, 50.3 MHz and 100.6 MHz, for the methylene group of polyisobutylene (PIB) in 70% PIB and 30% head-to-head poly(propylene) (hhPP) sample. The fitting curve of $\eta = 1$ is calculated based on the fitting parameters reported in Ref. [6], a traditional theoretical result.

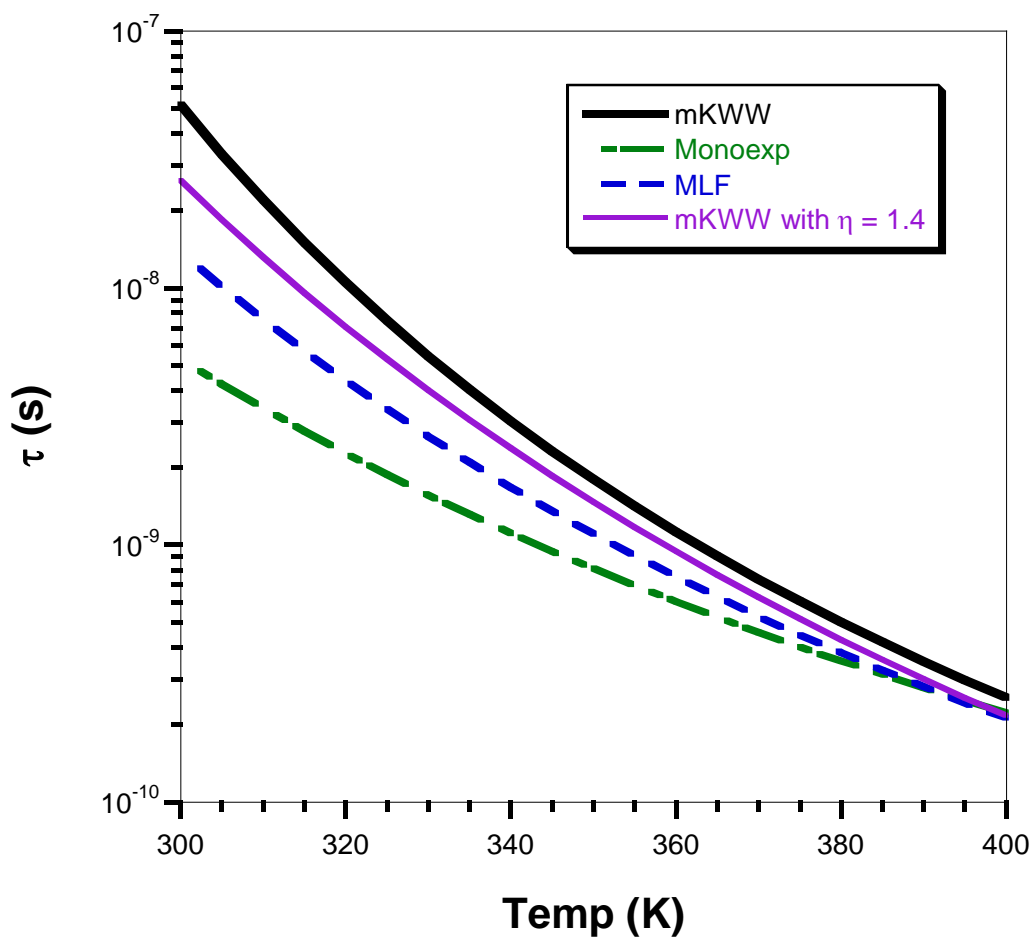

**Figure 3.** Comparison of temperature dependence segmental correlation times calculated based on the fittings of experimental data for the methylene group of polyisobutylene (PIB) in 70% PIB and 30% head-to-head poly(propylene) (hhPP) sample. The experimental $^{13}C$ $T_1$ NMR data are reported in Ref. [6], which is measured at variable temperatures and two field frequencies, 50.3 MHz and 100.6 MHz. The modified KWW (mKWW) segmental time is taken from Ref. [6]. The monoexponential, the mKWW with $\eta = 1.4$, and the MLF segmental times are obtained in this work based on the fit parameters listed in Table 1.

From the fitting in Figures 1 and 2, all the relaxation time models based on the MLF, the monoexponential function, and the modified KWW function can fit the data. For the mono-exponential model, the coupling constant $\eta$ is 1.97, which is near 2, indicating a strong coupling; while for the MLF and modified KWW based models, $\eta$ are 1.54 and 1.40, respectively. Both the MLF-based model and modified KWW based model may be equivalent to multiple-exponential modes that may make the coupling have less influence. Because the resonance in the NMR relaxation occurs when $(\eta\omega\tau)^2$ is near 1. the lower coupling $\eta$ means a larger $\tau$ value, which implies that the $\tau$ obtained from the traditional model, corresponding to $\eta = 1$, should have a larger $\tau$ value and slower motion, which is the exact case as shown in Figure 3. For comparison, the average segmental times from the MLF-based model adopt the same expression $\frac{\tau}{\alpha}\Gamma\left(\frac{1}{\alpha}\right)$ that has been used extensively for KWW function [5,6]. The motion in the modified KWW function based on the traditional model is obviously slower than the results from the current work. A similar phenomenon has been found in the study of the dynamics for the poly(ethylene oxide) [PEO] in miscible blends with poly(methyl methacrylate) [PMMA][25]; the dynamics of PEO obtained by the NMR results based on the traditional NMR model are somewhat slower than that obtained by Quasi-Elastic Neutron Scattering.

The slower correlation time from the traditional modified KWW function may be further explained in the following: The spectra density $J_{mKWW}(\omega)$ in Eq. (42b) includes two parts: liberational motion $a_{lib}\frac{\tau_{lib}}{1+(\eta\omega)^2\tau_{lib}^2}$, and non-liberational motion $(1-a_{lib})\sum_i\rho_i\frac{\tau_i}{1+(\eta\omega)^2\tau_i^2}$. Because the liberational motion is the fast motion which is often set as 1ps [5-6], and the NMR frequency is usually smaller than 1 GHz, $\omega^2\tau_{lib}^2 < 10^{-6}$, $\frac{\tau_{lib}}{1+(\eta\omega)^2\tau_{lib}^2} \approx 10^{-12} \approx 0$. Therefore, the liberational motion makes almost no contribution to $J_{mKWW}(\omega)$ for the NMR relaxation rate, however, the non-liberational motion contribution to $J_{mKWW}(\omega)$ in Eq. (42b) is reduced by the coefficient $1-a_{lib}$, which implies the $\tau_i$ from the conventional theory based on the modified KWW function could become $\frac{1}{1-a_{lib}}$ larger than it should be. $a_{lib}$ has been used in the analysis of quite a few NMR relaxation experiments in polymer systems [6].

Compared to the traditional theory, the phase diffusion method provides a significantly different view of liberational motion in Eq. (42b). The non-liberational motion overcomes the energy barrier, while the liberational motion does not overcome the energy barrier, so it often is a fast motion vibrating inside a relatively small spatial region The phase jump of the liberational motion and the subsequent random motion should keep direction, therefore, from the view of random walk theory, the liberational and the subsequent non-liberational motions should are not two separate phase jumps, but just a single phase jump with a slightly increased waiting time. Consequently, the NMR relaxation could not directly detect the liberational motion but see the liberational motion as part of the jump of non-liberational motion. Although both the modified KWW function with $\eta = 1$ and $\eta = 1.4$ can fit the experimental data taken from Ref. [6], the parameter $a_{lib}$ is different. The traditional theory with $\eta = 1$, $a_{lib} = 0.26$ may get arbitraly slower charateric time. While the phase-time coupling yields $\eta = 1.4$, $a_{lib} = 0$, which implies the librational motion may not be able to detect in NMR relaxation experiments.

The spectral density from a monoexponential function based on conventional theory cannot fit this data. However, it is interesting that the monoexponential with a coupling constant, and modified KWW and MLF based relaxation times with the coupling constant $\eta$ can fit the experimental data. This may be due to the following: The specific experimental data obtained from Ref. [6] shown in Figure 1 may not be sensitive to the difference between the two different types of time distributions; additionally, the relaxation time Eq. (41a) has already included three different angular frequencies.

The current paper neglects the effect that the amplitude of $\omega_0$ could obey a distribution. It needs further research to understand how the distribution of $\omega_0$ affects the coupling effect. More effort is needed to apply this model to fit more experimental data. Unlike the phase-time coupling naturally occurring in the NMR process, the relaxation phenomena in other techniques, such as dielectric relaxation [6] should not observe similar coupling behaviors. Comparing to the results from other techniques may improve our understanding of the model. To my best knowledge, the phase and time coupling concept has not been considered for NMR relaxation; further efforts are needed to better understand the phase and time coupling effect and to improve the current method.

# APPENDIX A. QUANTUM MECHANICS DESCRIBING OF NMR RELAXATION BY SECOND ORDER PERTURBATION

Under the external magnetic field $H_0$ and the randomly fluctuating field $H_1(t)$, the density operator $\sigma$ of the spin system evolves according to [1-3]

$$\frac{d\sigma^*}{dt} = -i[H_1^*(t), \sigma^*], \tag{A.1}$$

where $\sigma^* = e^{iH_0 t}\sigma e^{-iH_0 t}$, and $H_1^*(t)$ can be expressed as [1-3]

$$H_1^*(t) = e^{iH_0 t} H_1(t) e^{-iH_0 t} = \sum_q F^{(q)} A^{(q)} = \sum_{pq} F^{(q)} A_p^{(q)} e^{i\omega_p^{(q)} t}, \tag{A.2}$$

where $F^{(q)}$ are the lattice operators and $A_p^{(q)}$ are the spin operators. Substituted Eq. (A.2) into Eq. (A.1) gives

$$\frac{d\sigma^*}{dt} = -i \sum_{p,q} F^{(q)} \left[ A_p^{(q)} e^{i\omega_p^{(q)} t}, \sigma^* \right]. \tag{A.3}$$

In Eq. (A.3), the starting relative position of the Hamiltonian affects $F^{(q)}$'s amplitude and determines its value to be positive or negative. $A^{(q)} e^{i\omega_p^{(q)} t}$ could be seen as a rotating vector with frequency $\omega_p^{(q)}$, which drives the evolution of the state vector or the density operator $\sigma^*$ at an effective frequency $|F^{(q)} A^{(q)}|$. In the ensemble of all spins, or as the average of the individual spin in a whole random walk process, only one component of the observable vector is measured in NMR relaxation experiments, and thus the effective phase change during a jump time interval is its real part $\int_0^{t_i} \omega_0 \cos(\omega t) dt$, where $\omega_0 = |H_1(t)| = |F^{(q)} A^{(q)}|$.

While the traditional second-order perturbation theory does not consider the instantaneous projection $\int_0^{t_i} \omega_0 \cos(\omega t) dt$. Instead, various approximations are employed to obtain the approximated $\sigma^*(t)$. The strategy of the traditional method is briefly described in the following:

Performing the integration on both sides of Eq. (A.1), we have

$$\sigma^*(t) = \sigma^*(0) - i \int_0^t dt' [H_1^*(t), \sigma^*(t)]. \tag{A.4}$$

$\sigma^*(t)$ in the right-hand side of Eq. (A.4) can be approximately replaced by

$$\sigma^*(t) \approx \sigma^*(0) - i \int_0^t dt' [H_1^*(t), \sigma^*(0)], \tag{A.5}$$

to give

$$\sigma^*(t) \approx \sigma^*(0) - i \left\{ \int_0^t dt' [H_1^*(t), \sigma^*(0)] - i \int_0^t dt' \int_0^{t'} dt'' [H_1^*(t'), [H_1^*(t''), \sigma^*(0)]] \right\}. \tag{A.6}$$

Performing derivation on both sides of Eq. (A.6) gives us

$$\frac{d}{dt}\sigma^*(t) = -i[H_1^*(t), \sigma^*(0)] - \int_0^{t'} dt'' [H_1^*(t'), [H_1^*(t''), \sigma^*(0)]]. \tag{A.7}$$

By performing ensemble average on Eq. (A.7), $\overline{H_1^*(t)} = 0$, and replacing $t'$ with $t$, and setting $t'' = t - \tau$, we have

$$\frac{d}{dt}\sigma^*(t) = -\int_0^t d\tau \overline{[H_1^*(t), [H_1^*(t - \tau), \sigma^*(0)]]}. \tag{A.8}$$

To obtain the relaxation expression, it needs further approximations; the integral region from 0 to $t$ in Eq. (A.8) is approximately extended from 0 to infinity, and $\sigma^*(0)$ is approximately replaced with $\sigma^*(t)$. Then we have

$$\frac{d}{dt}\sigma^*(t) \approx -\int_0^\infty d\tau \overline{[H_1^*(t), [H_1^*(t - \tau), \sigma^*(t)]]}. \tag{A.9}$$

Substituted Eq. (A.2) into Eq. (A.9), we have

$$\frac{d}{dt}\sigma^*(t) = -\sum_{p,p',q,q'} e^{i\left(\omega_p^{(q)} + \omega_{p'}^{(q')}\right)t} \left[ A_{p'}^{(q')}, [A_p^{(q)}, \sigma^*(t)] \right] \int_0^\infty d\tau \overline{F^{(q')}(t) F^{(q)}(t - \tau)} e^{i\omega_p^{(-q)} \tau}. \tag{A.10}$$

By assuming $q = -q'$, $e^{i\left(\omega_p^{(q)} + \omega_{p'}^{(q')}\right)t} = 1$, and neglecting the non-secular terms, Eq. (A.10) reduces to

$$\frac{d}{dt}\sigma^*(t) \approx -\sum_{p,q} \left[ A_p^{(-q)}, [A_p^{(q)}, \sigma^*(t)] \right] \int_0^\infty d\tau \overline{F^{(-q)}(t) F^{(q)}(t - \tau)} e^{i\omega_p^{(-q)} \tau}. \tag{A.11}$$

which is the fundamental equation for the traditional NMR relaxation theory. The time correlation function $G(t)$ for NMR relaxation is

$$G(t) \propto \overline{F^{(q)}(t-\tau)\, F^{(-q)}(t)} \equiv \overline{F^{(q)}(t)\, F^{(-q)}(t+\tau)}. \tag{A.12}$$

Almost all four approximations used in the above derivation are unnecessary in the phased diffusion method. The phase diffusion method assumes that the NMR observable, such as angular momentum, undergoes a random phase walk, which could be treated by the phase diffusion or random walk method. For a random field induced by a Hamiltonian in a static frame or without considering phase-time coupling, the phase diffusion method gives the same NMR relaxation rate as that obtained by the second-order perturbation theory mentioned above. When the phase-time coupling is considered, the effective angular frequency is modified by $\eta\omega$.